\documentclass{aa}  

\usepackage{graphicx}
\usepackage{txfonts}
\usepackage{natbib}
\bibpunct[; ]{(}{)}{;}{a}{}{,}
\newcommand{\ha}{$H_\alpha$ }

\begin{document}

   \title{Orbital and physical parameters of eclipsing binaries from the All-Sky Automated Survey catalogue}

   \subtitle{VI. AK~Fornacis -- a rare, bright K-type eclipsing binary\thanks{Based 
on observations collected at the European Southern Observatory, Chile under programmes 088.D-0080, 089.D-0097, 089.C-0415, 090.C-0280 and 090.D-0061, and through CNTAC proposals CN-2012A-21 and CN-2013A-93.}}

\author{K. G. He\l miniak \inst{1,2,3}
	\and
	R. Brahm\inst{1}
	\and
	M. Ratajczak\inst{2}
	\and
	N. Espinoza\inst{1}
	\and
	A. Jord\'an\inst{4,1}
	\and
	M. Konacki\inst{2,5}
	\and
	M. Rabus\inst{1,6}
          }

   \institute{
Pontificia Universidad Cat\'{o}lica de Chile, Av. Vicu\~{n}a Mackenna 4860, 7820436 Macul, Santiago, Chile\\
              \email{xysiek,rbrahm,nsespino,ajordan,mrabus@astro.puc.cl}
         \and
Nicolaus Copernicus Astronomical Center, Department of Astrophysics, ul. Rabia\'{n}ska 8, 87-100 Toru\'{n}, Poland\\
             \email{xysiek,milena,maciej@ncac.torun.pl}
	\and
Subaru Telescope, National Astronomical Observatory of Japan, 650 North Aohoku Place, 
Hilo, HI 96720, USA
	\and
Millennium Institute of Astrophysics, Av.\ Vicu\~na Mackenna
4860, 7820436 Macul, Santiago, Chile
	\and
Astronomical  Observatory, A. Mickiewicz University, ul. S\l oneczna 36, 60-286 Pozna\'n, Poland
	\and
Max-Planck-Institut f\"ur Astronomie, K\"onigstuhl 17, D-69117 Heidelberg, Germany
             }

   \date{Received ...; accepted ...}

 
  \abstract
 {} 
   {We present the results of the combined photometric and 
spectroscopic analysis of a bright ($V=9.14$), nearby 
($d=31$ pc), late-type detached eclipsing binary 
\object{AK~Fornacis}. 
This $P=3.981$ d system has not been previously
recognised as a double-lined spectroscopic binary, and 
this is the first full physical model of this unique target.}
   {With the FEROS, CORALIE and HARPS spectrographs we collected a number of 
high-resolution spectra in order to calculate radial velocities 
of both components of the binary. Measurements were done with our 
own disentangling procedure and
the TODCOR technique, and were later combined with the photometry from the
ASAS and SuperWASP archives. We also performed an atmospheric
analysis of the component spectra with the Spectroscopy Made Easy (SME) package.
}
   {Our analysis shows that AK~For consists of two active, cool dwarfs 
having masses of $M_1=0.6958 \pm 0.0010$ and $M_2=0.6355 \pm 0.0007$ M$_\odot$ 
and radii of $R_1=0.687 \pm 0.020$ and $R_2=0.609 \pm 0.016$ R$_\odot$, 
slightly less metal abundant than the Sun. 
Parameters of both components are well reproduced by the models.
}
   {AK~For is the brightest system among the known eclipsing 
binaries with K or M type stars. Its orbital period is 
one of the longest and rotational velocities one of the 
lowest, which allows us to obtain very precise radial velocity measurements. The 
precision in physical parameters we obtained places AK~For among 
the binaries with the best mass measurements in the literature. 
It also fills the gap in our knowledge of stars in the range of
0.5--0.8 M$_\odot$, and between short and long-period systems.
All this makes AK~For a unique benchmark for understanding the properties of 
low-mass stars.}

   \keywords{binaries: eclipsing -- binaries: spectroscopic -- stars: fundamental parameters -- stars: late-type -- stars: individual: AK~Fornacis
               }

   \maketitle
%

\section{Introduction}

During the last years we have witnessed a remarkable improvement in the 
study of lower main-sequence stars, mainly by the increasing number of 
eclipsing binaries found to contain K and M type dwarfs. The interest in 
those objects has increased recently with the new exoplanet search surveys 
dedicated especially to them, such as
the WFCAM Transit Survey\footnote{\texttt{http://www.ast.cam.ac.uk/$\sim$sth/wts/index.html}}. 
The knowledge of the host star parameters, especially mass and radius, 
is necessary to constrain the characteristics of the planet \citep{dem11}.
Another justification for studying late type stars are the still
unexplained discrepancies between the models and observations. We still
do not fully understand why in most cases, sometimes surprisingly, the stars are 
larger and cooler than predicted \citep[e.g.][]{lac77,pop97,tor02,mor09,irw11}, 
while in others the same characteristics are nicely reproduced 
\citep{tho10,fei11,hel11a}. Unfortunately, most of newly discovered systems
tend to be faint, and have large rotational velocities due to short periods 
($P < 3$~d), thus their properties are difficult to measure accurately.
Only a handful of them have their masses and radii determined with the accuracy 
better than 3\% (with systematics taken into account properly), thus allowing for 
meaningful tests of the stellar models \citep{tor10}. 
Each new system is highly-valuable, and in
this paper we present the analysis of the brightest one known to date, 
which also has one of the most accurate mass measurements among the 
eclipsing binaries untill now.

\begin{table}
\caption{Literature and catalogue data for AK~For}\label{tab_lit}
\centering
\begin{tabular}{lcc}
\hline  \hline
Parameter & Value & Ref. \\
\hline
$\alpha_{ICRS}$	&  03:29:22.87422 & 1 \\
$\delta_{ICRS}$	& -24:06:03.0926 & 1 \\
$\pi$ [mas]	& 32.19(1.13) & 1 \\
$\mu_\alpha$ [mas/yr] & 220.58(95) & 1 \\
$\mu_\delta$ [mas/yr]& 99.53(1.66) & 1 \\
$B$ [mag]	& 10.46	& 2	\\
$V_{Mer}$ [mag]	&  9.36	& 2	\\
$V_{ASAS}$ [mag]	& 9.135 & 3	\\
$J$ [mag]	& 7.031(21)& 4	\\
$H$ [mag]	& 6.510(53)& 4	\\
$K$ [mag]	& 6.262(33)& 4	\\
Sp. Type	& K3~V~k~Fe+0.4& 5 \\
$P_{orb}$ [d]		& 3.981	& 6 \\
\hline
\end{tabular}
\tablefoot{
1: {\it Hipparcos} \citep{vLe07};
2: \citet{mer86};
3: our fit to ASAS data;
4: 2MASS \citep{skr06};
5: \citet{gra06};
6: \citet{ote03}.}
\end{table}

The variable nature of AK~Fornacis (HD~21703, HIP~16247, ASAS~J032923-2406.1;
hereafter \object{AK~For}) was discovered by the \textit{Hipparcos} mission, and the star
was classified as a probably eclipsing in the 74th Special Name-list of Variable Stars 
\citep{kaz99}. Before that it was only known as a K-type star 
\citep{upg72} and an X-ray source \citep{gio90}.
\citet{ote03} gave the first period estimation -- 3.981 d -- on the basis of
\textit{Hipparcos} and the All-Sky Automated Survey \citep[ASAS;][]{poj02}
data available that time, however
AK~For is not listed in the ASAS Catalogue of Variable Stars. The only spectroscopic
study was done in low resolution by \citet{gra06}, who classified AK~For as a 
K3~V, ``chromospherically very active'' star. Finally, \citet{bai11} estimated
the effective temperature and extinction towards the star on the basis of 
multi-band photometry, but his method is not suitable for binaries, so both values
are overestimated (5350~K, 1.28~mag). The summary of catalogue and literature
data is presented in Table \ref{tab_lit}. Note the discrepancy between ASAS and
Mermilliod's (\citeyear{mer86}) values of the $V$ magnitude. The system's $T_{eff}$ 
calculated from $V-K$ colours and calibrations by \citet{wor11} is $\sim$4100~K for
\citet{mer86} and $\sim$4350~K for ASAS.

\section{Observations} 
\subsection{Spectroscopy}
With the Fiber-fed Extended Range Optical Spectrograph \citep[FEROS;][]{kau99} 
we collected twelve high-resolution ($R\sim40\,000$) spectra in November 2011,
September 2012, and March 2013. They were supplemented with six $R\sim70\,000$ spectra 
taken in February 2013 with the CORALIE spectrograph, attached to the 1.2-m Euler 
telescope in La Silla, and two $R\sim115\,000$ spectra from the High Accuracy 
Radial velocity Planet Searcher \citep[HARPS;][]{may03} obtained in quadrature on 
September 09, 2012. 
FEROS and CORALIE data were reduced with a dedicated Python-based pipeline
\citep[for a description see:][]{jor14},
which was initially built for CORALIE data reduction, 
but was modified to deal with FEROS data. HARPS spectra were reduced on-site 
with the available Data Reduction Software (DRS).

\subsection{Photometry}
The $V$-band photometry of AK~For, publicly available from the ASAS 
Catalogue\footnote{\texttt{http://www.astrouw.edu.pl/asas/?page=aasc}},
spans from November 2000 to December 2009, and contains 883 good quality 
points (flagged ``A'' in the original data). The time span of the public 
Wide-Angle Search for Planets \citep[SuperWASP;][]{pol06}
data\footnote{\texttt{http://exoplanetarchive.ipac.caltech.edu/applications\\/ExoTables/search.html?dataset=superwasptimeseries}} 
is much shorter -- from August 2006 till February 2010, but with only a 
few, practically useless points after Feb. 2008. Due to the orbital period very 
close to 4 days, the eclipses are not always observable in the night, so
in the case of SuperWASP we initially selected only 2007-2008 data, where 
eclipses were visible. We later limited ourselves to 2709 data points with 
individual errors smaller than 0.03 mag. This threshold level was a
result of the further analysis, in which we wanted to keep as many good
quality data points as possible (especially in eclipses), while at the same time 
minimizing the systematics in the resulting radii. We also noted that depths of 
single minima vary slightly, but it was impossible to conclude if 
it is a real effect (due to spots for example) or an uncorrected 
systematic. AK~For is known to be an active system but no consistent
out-of-eclipse modulation was found in the SuperWASP photometry.

To check for the spot-originated brightness variations, we performed $V$-band
observations with the PISCO telescope, attached to the main tube of 
the 1.2-m Euler telescope in La Silla. PISCO is a 31.75 cm (12.5 in) 
aperture reflector by RC Optical Systems, with a Finger Lakes Instrumentation 
(FLI) Proline-09000 camera, equipped with a $3056 \times 3056$ pixel Kodak 
KAF-09000 CCD. With the pixel scale of 0.862~''/pix, the field 
of view is $44'' \times 44''$, strongly vignetted in the corners. 
It contains a set of standard Johnson-Cousins $U,B,V,R,I$ filters, 
supplemented by near-IR $Z$ and $Y$. It works together with the CORALIE 
spectrograph allowing for a simultaneous photometry and spectroscopy, 
taking exposures while CORALIE's shutter is open. With PISCO we secured 58 
images in the $V$ band, with exposure times of 18 sec, taken exactly in the 
same time as CORALIE spectra from nights February 24, 25 and 26, 2013. 
Standard CCD reduction steps, utilizing bias and sky-flat images, were 
performed in IRAF, as well as a simple aperture photometry under the task 
\texttt{apphot}. AK~For was by far the brightest star in the field, so 
the selection of comparison stars was difficult. We inspected seven other 
stars, and found two that did not show any significant brightness
variations but unfortunately are about 2 mag fainter than AK~For.
We only wanted to perform a
relative photometry, in order to detect any variability, so we did not transform 
PISCO magnitudes to the standard Johnson's system, nor we observed photometric 
standard stars.

\section{Analysis}

\subsection{Radial velocities and the orbital solution}

   \begin{figure}
   \centering
   \includegraphics[width=\columnwidth]{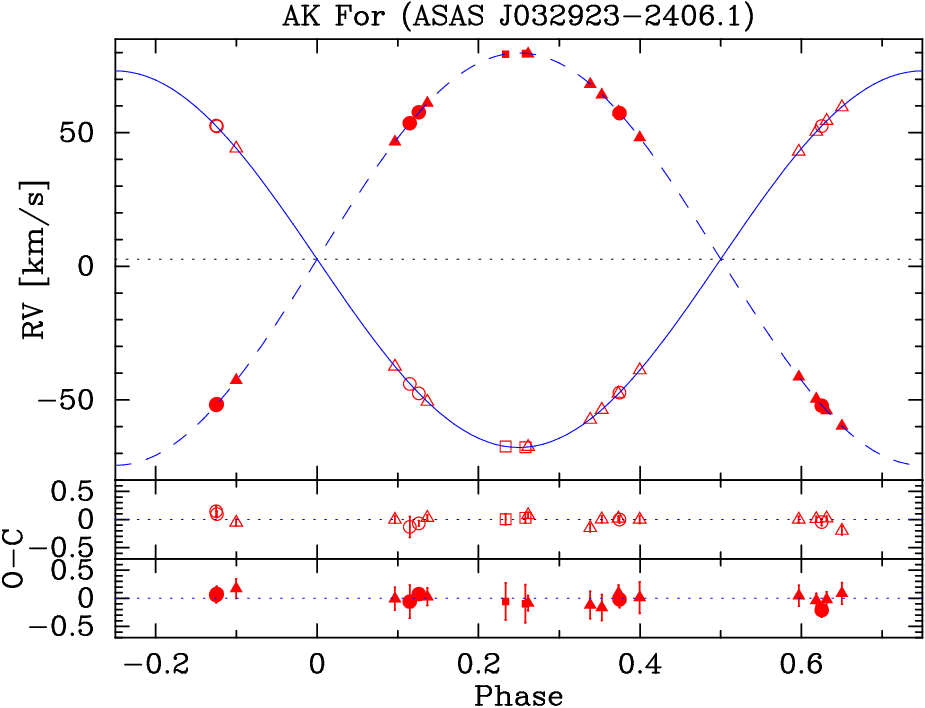}
   \caption{Radial velocities and orbital solution for AK~For. Solid line 
and empty symbols refer to the primary, dashed line and filled symbols 
to the secondary component. Triangles mark the FEROS, circles mark the CORALIE 
and squares the HARPS data. The systemic velocity is marked with the dotted line. 
The residuals are plotted below together with the (scaled) individual errors.
The $rms$ of the orbital fit is 82 and 96 m/s for the primary and secondary respectively.
}\label{fig_rv}
   \end{figure}

   \begin{figure}
   \centering
   \includegraphics[width=\columnwidth]{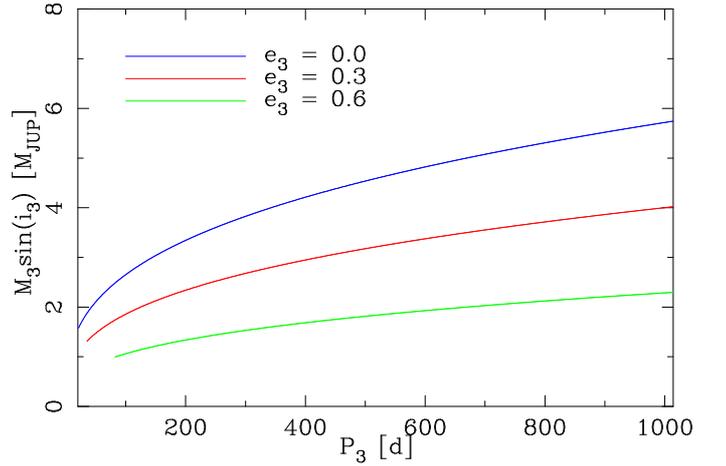}
   \caption{Third body detection $M_3\sin(i_3)$ limits estimated from
our RV measurements, for several values of outer orbit's eccentricities. 
We should be able to detect bodies with masses above these limits.}\label{fig_3rd}
   \end{figure}

\begin{table*}
\caption{Individual RV measurements, errors and residuals (all in km/s) of the components of AK~For, together with the orbital phase, exposure times (in seconds) and S/N of the spectra. 2F marks the MPG 2.2-m/FEROS, 3H the ESO 3.6-m/HARPS, and EC the 1.2-m Euler/CORALIE data (with ``p'' meaning simultaneous observations with PISCO).}\label{tab_rv}
\centering
\begin{tabular}{lrrrrrrcrrc}
\hline  \hline
JD-2450000 & $v_1$ & $\sigma_1$ & $(O-C)_1$ & $v_2$ & $\sigma_2$ & $(O-C)_2$ & Phase & $T_{exp}$ & S/N & Tel./Sp. \\
\hline
5876.681435 & -37.483 & 0.062 & -0.003 &  46.441 & 0.202 & -0.008 & 0.0965 &  450 &  90 & 2F \\
5876.841349 & -50.637 & 0.049 &  0.031 &  60.928 & 0.155 &  0.026 & 0.1366 &  900 & 125 & 2F \\
5877.644573 & -57.370 & 0.067 & -0.149 &  67.959 & 0.242 & -0.122 & 0.3384 &  750 & 120 & 2F \\
5877.701489 & -53.645 & 0.061 &  0.003 &  64.000 & 0.231 & -0.165 & 0.3527 & 1200 &  90 & 2F \\
5877.783808 & -47.668 & 0.038 &  0.018 &  57.725 & 0.139 &  0.092 & 0.3734 & 1200 &  95 & 2F \\
5878.673272 &  42.913 & 0.056 & -0.001 & -41.535 & 0.185 &  0.043 & 0.5968 &  600 &  95 & 2F \\
5878.759453 &  50.399 & 0.043 &  0.010 & -49.798 & 0.126 & -0.041 & 0.6184 &  450 &  90 & 2F \\
5878.810493 &  54.428 & 0.039 &  0.021 & -54.178 & 0.137 & -0.024 & 0.6313 &  480 & 105 & 2F \\
6179.782690 & -67.426 & 0.095 &  0.004 &  79.214 & 0.329 & -0.056 & 0.2336 &  900 &  85 & 3H \\
6179.879929 & -67.687 & 0.089 &  0.027 &  79.484 & 0.339 & -0.097 & 0.2580 &  780 &  90 & 3H \\
6195.817064 & -67.545 & 0.048 &  0.080 &  79.397 & 0.134 & -0.086 & 0.2613 &  480 & 120 & 2F \\
6345.559577 &  52.316 & 0.054 &  0.094 & -51.594 & 0.136 &  0.075 & 0.8757 &  420 &  35 & EC \\
6346.511053 & -44.029 & 0.184 & -0.132 &  53.513 & 0.294 & -0.061 & 0.1147 &  600 &  25 & EC \\
6347.545693 & -47.323 & 0.060 & -0.001 &  57.305 & 0.144 & -0.021 & 0.3746 &  600 &  30 & EC \\
6348.542794 &  52.439 & 0.060 & -0.046 & -52.171 & 0.123 & -0.212 & 0.6251 &  480 &  30 & ECp \\
6349.537454 &  52.613 & 0.091 &  0.146 & -51.882 & 0.134 &  0.055 & 0.8749 &  420 &  25 & ECp \\
6350.536805 & -47.586 & 0.046 & -0.069 &  57.613 & 0.093 &  0.072 & 0.1259 &  600 &  30 & ECp \\
6376.509645 &  59.529 & 0.075 & -0.198 & -59.889 & 0.191 &  0.086 & 0.6502 &  600 &  90 & 2F \\
6377.503283 &  44.095 & 0.062 & -0.054 & -42.758 & 0.173 &  0.172 & 0.8997 &  600 & 135 & 2F \\
6383.474455 & -38.866 & 0.068 & -0.001 &  47.980 & 0.279 &  0.012 & 0.3997 &  600 &  75 & 2F \\
\hline
\end{tabular}
\end{table*}

Radial velocities (RVs) were initially calculated
with our implementation of the two dimensional cross-correlation technique 
\citep[TODCOR;][]{zuc94} with synthetic spectra used as templates.
Due to the poor S/N in the blue part, limits implemented to the CORALIE 
and FEROS reduction pipelines (optimized for high-precision RVs), 
and limitations of the synthetic templates, we used the following 
wavelength ranges: 4135--6500 \AA\ for FEROS, 4400--6500 \AA\ 
for CORALIE, and 3800--6500 \AA\ for HARPS.
One of the TODCOR's features is that it also gives the most probable 
flux ratio of the two components. We found it to be 0.423 and that 
it is pretty stable along the $V$ and SuperWASP bands.
Later we used the FEROS spectra to perform disentangling, as 
described in \citet{kon10}, and obtain separate spectra of the two 
components. We used them to perform the spectral analysis, and for final 
RV measurements. For the FEROS data we followed the procedure of 
least-squares fitting described in \citet{kon10}, and for CORALIE and 
HARPS we used the disentangled spectra as templates in TODCOR. 

The RV measurements were analysed with a simple procedure, which fits a 
double-keplerian orbit using the Levenberg-Marquartd algorithm. As free 
parameters we set the orbital period $P$, the time of primary conjunction $T_0$, 
corresponding in this case to the primary (deeper) eclipse, the velocity 
semi-amplitudes $K_{1,2}$, primary's systemic velocity $\gamma_1$, the difference
in components systemic velocities $\gamma_2-\gamma_1$, and the difference between 
FEROS and CORALIE zero-points 2F--EC, measured separately for each component. 
We assumed that HARPS has the same zero-point as FEROS, as we did not have 
enough HARPS data to securely constrain this difference. HARPS measurements 
do not outlay significantly from the final solution, so we find this assumption 
correct. Initially, we have also fitted for the eccentricity $e$ and the argument 
of the periastron $\omega$ but neither in this step, nor in any other, we 
found the orbit eccentric. Thus we kept $e=0$ during the whole process. Finally, 
we ended up with 40 data points (20 for each component) and 8 parameters.

To estimate reliable parameter uncertainties, with systematics accounted for,
we run 10000 bootstrap iterations. 
All the RV measurements, together with their
uncertainties, residuals from the model RV curve, orbital phases, exposure
times and S/N around 5500\AA, are shown in the Table~\ref{tab_rv}.
The resulting orbital parameters are presented in Table~\ref{tab_par}.

The precision of RVs we have reached is good enough to look for massive 
circumbinary planets around the system. If we assume that we can detect 
any RV variation of the semi-amplitude of at least 96 m/s -- the $rms$ of 
the secondary -- we can calculate the mas of a body that could produce such 
a variation. We show those calculations in Figure \ref{fig_3rd} as a function 
of the outer orbit's period and for three values of the outer orbit's eccentricity: 
0.0, 0.3 and 0.6. The value of the 96~m/s was assumed to be the RV semi-amplitude. 
The lower limit of the period is 21~d (the major semi-axis of 34.83~R$_\odot$) 
which refers to the shortest stable circular orbit \citep{hol99}. For eccentric
orbits the limits are terminated at the shortest periods having the distance of
periastron larger than 34.83~R$_\odot$ (around 35 and 82~d). The upper
limit of the period is 1014~d, twice the time span of our data. One can see that
our radial velocities are precise enough to detect massive planets on stable
orbits around AK~For. If there was a third body orbiting within this range of
periods, it must have the mass lower than the limit.

\subsection{Atmospheric parameters}

   \begin{figure*}
   \centering
   \includegraphics[width=0.9\textwidth]{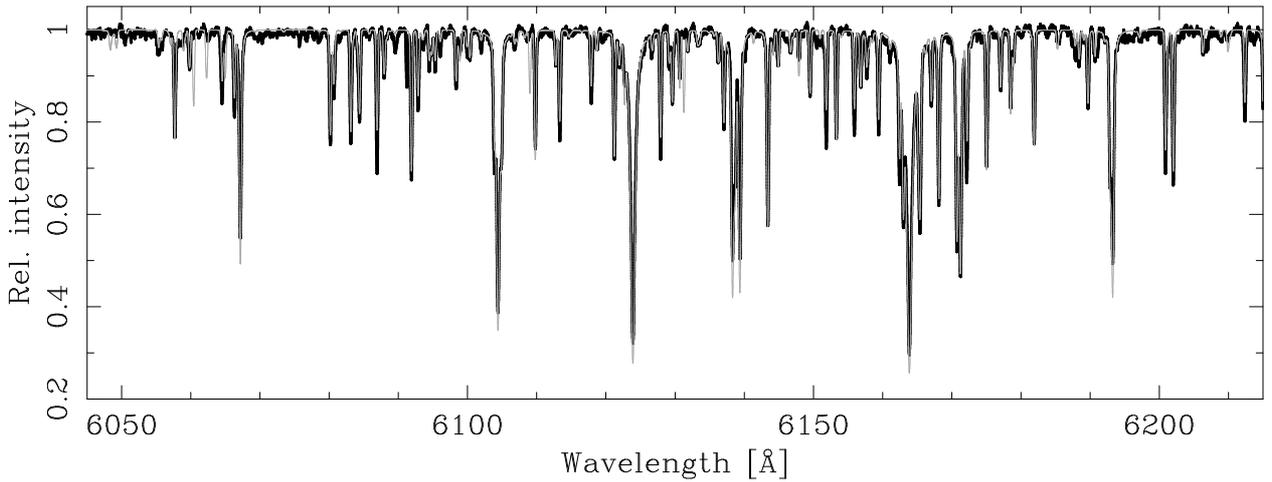}
   \caption{Portion of the SME best fit (thin grey line) to the disentangled spectrum of the primary (thick black line).}\label{fig_sme}
   \end{figure*}

We performed a spectral atmospheric analysis using the Spectroscopy Made Easy
package \citep[hereafter SME;][]{val96}. 
The disentangled spectra were first scaled with the flux ratio obtained 
from TODCOR (0.423). We've also run SME with other flux ratios, but the 
resulting temperatures led to further inconsistencies, in distances for example.

Following the work of \citet{val05}, in order to obtain a continuum placement 
and derived parameters more accurate, we have chosen seven FEROS orders 
between 5317 and 6397~\AA, each of them was analysed separately. 
We used the list of atomic lines from Vienna Atomic 
Line Database \citep[VALD;][]{pis95,kup99} generated for the Sun for the 
initial values described by \citet{val05}. We also adopted atmosphere 
models of \citet{kur93}. We set the $\log(g)$ for each component to the
values obtained with JKTABSDIM (see: Tab. \ref{tab_par} and next 
Section), micro- and macro-turbulence velocities to 1 and 3 km/s respectively,
while $T_{eff}$, $[M/H]$ and $v \sin(i)$ were fitted for every order.
As starting values of temperatures we used 4500 and 4150 K for the primary 
and secondary respectively. For each order we also started with three different 
values of $[M/H]$: -1, -0.5 and 0. From all SME runs we took only those,
for which the resulting $v \sin(i)$ was close to the value obtained from
JKTEBOP (see next Section). Of the 21 SME runs for each component, 
only 13 and 9 was successful for the primary and secondary respectively.
Example of a successful fit to one of the orders of the primary's spectrum
is presented in Figure~\ref{fig_sme}.

For the adopted values of effective temperatures, metallicities
and projected rotational velocities we took the averages, and standard 
deviations as their uncertainties. Those values are given in 
Table \ref{tab_par}. The ratio of velocities is in agreement with the
ratio of radii found in the light curve fitting (see next Section), 
but the velocities themselves are larger than predicted by the final 
solution and tidal locking. This could have been caused by underestimation
of the assumed micro- or macro-turbulence velocity. What is surprising
we found two significantly different values of $[M/H]$: -0.3 and -0.1
for the primary and secondary respectively, both with uncertainties of
0.05 dex. We found that this difference was not dependent on the 
starting values of any parameter, including the flux ratio, and was 
not correlated with the resulting temperatures. We need to note however, 
that the FEROS spectra were not corrected for the scattered light, which 
might have influenced the SME analysis, and that the S/N of the 
secondary's spectrum is significantly lower than the primary's. We thus 
adopt a conservative value of $[M/H]=-0.2\pm0.1$~dex.

\subsection{Light curve solution}

   \begin{figure*}
   \centering
\includegraphics[width=0.8\textwidth]{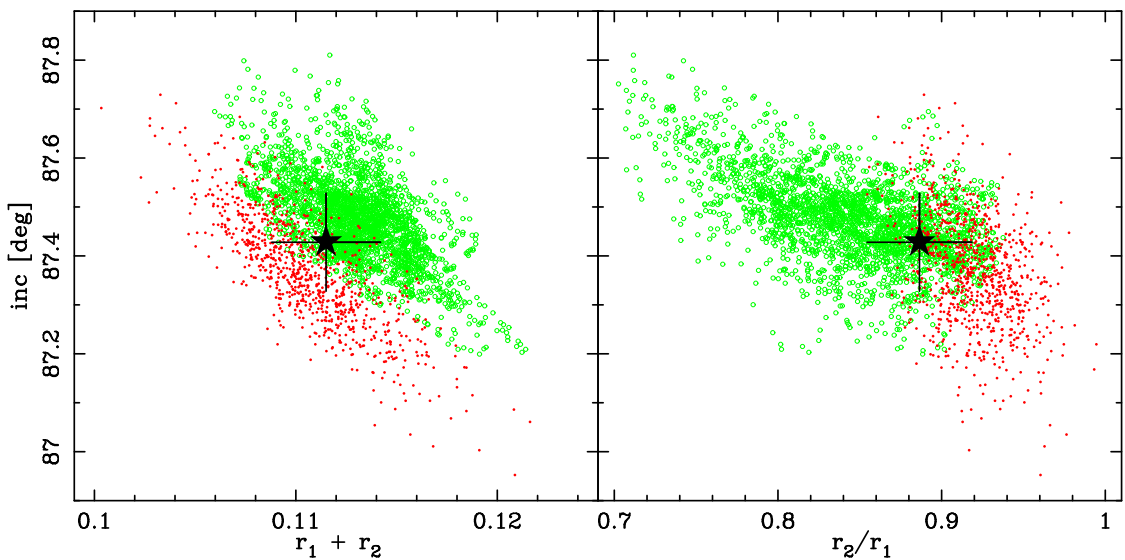}
   \caption{Results of the residual-shifts analysis performed with JKTEBOP 
on the ASAS (red) and SuperWASP data (green). Plots present the distribution
of consecutive solutions on the $r_1 + r_2$ vs. $i$ (left) and 
$k=r_2/r_1$ vs. $i$ (right) panels. Black stars with error bars correspond 
to the adopted values with their $1\sigma$ uncertainties.}
              \label{fig_inc_rr}
    \end{figure*}

   \begin{figure*}
   \centering
\includegraphics[width=\textwidth]{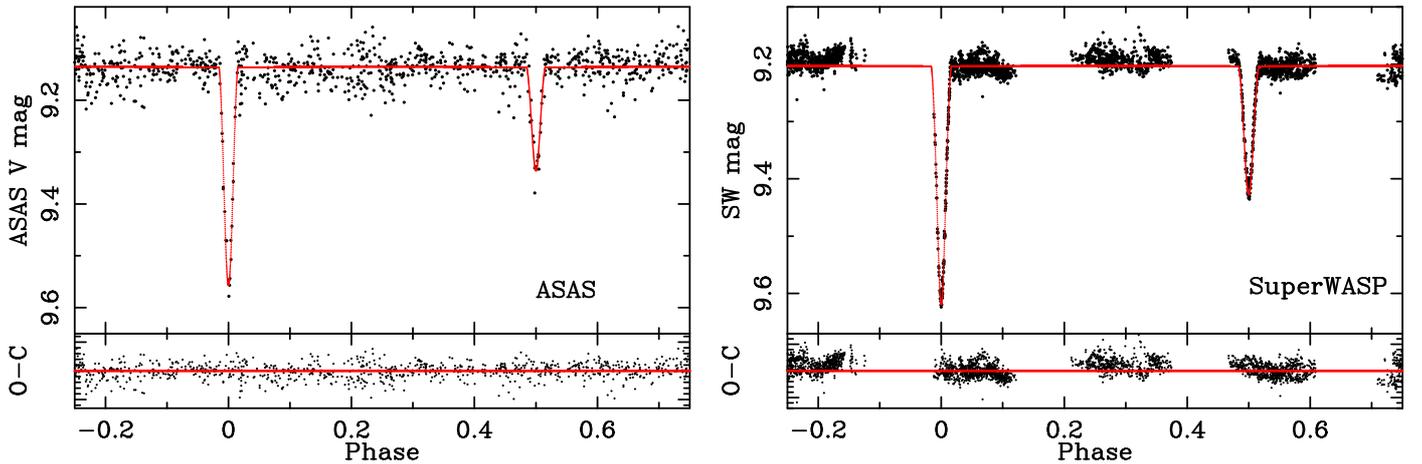}
   \caption{Phase-folded light curves of AK~For from ASAS (left) and SuperWASP (right) together with the
best-fitting model and residuals.}
              \label{fig_lc}
    \end{figure*}

\begin{table}
\caption{\label{tab_par}
Orbital and physical parameters of AK~For.
}
\centering
\begin{tabular}{lcccc}
\hline\hline
{\bf AK For}& \multicolumn{2}{c}{Primary} & \multicolumn{2}{c}{Secondary}\\
Parameter & Value & $\pm$ & Value & $\pm$\\
\hline
&\multicolumn{4}{c}{\it Radial velocity fit} \\
$P$ [d]		& \multicolumn{4}{c}{$3.9809913 \pm 0.0000045$}\\
$T_0$ [JD]	& \multicolumn{4}{c}{$2451903.2681 \pm 0.0045$}\\
$K$ [km/s]	& 70.471& 0.031 & 77.136& 0.051 \\
$\gamma$ [km/s] & 2.667 & 0.028 & 2.516 & 0.115 \\
$M\sin^3(i)$ [M$_\sun]$	& 0.6937 & 0.0010& 0.6336& 0.0008\\
2F--EC [km/s]	&-0.037 & 0.032 & 0.133 & 0.073 \\
2F--3H [km/s]	&  0.0  &  ---  &  0.0  &  ---  \\
$rms$ [km/s]	& \multicolumn{2}{c}{0.082} & \multicolumn{2}{c}{0.096}\\
&\multicolumn{4}{c}{\it SME spectral analysis} \\
$T_{eff}$ [K]	&  4690 &  100  &  4390 &  150 \\
$[M/H]$		& -0.29 & 0.05  & -0.10 & -0.05 \\
$v\sin(i)$ [km/s] & 9.52 & 0.22 &  8.27 & 0.42 \\ 
&\multicolumn{4}{c}{\it Light curve fit\tablefootmark{a}} \\
$i$ [$^\circ$]	& \multicolumn{4}{c}{$87.428 \pm 0.080$}\\
$r$ $[R/a]$	& 0.0591& 0.0017& 0.0524& 0.0014\\
$J_{ASAS}$	&\multicolumn{4}{c}{$0.528 \pm 0.028$}\\
$J_{SW}$	&\multicolumn{4}{c}{$0.592 \pm 0.030$}\\
%
&\multicolumn{4}{c}{\it Adopted physical parameters} \\
$a$ [R$_\sun$]	& \multicolumn{4}{c}{$11.6318\pm 0.0047$} \\
$M$ [M$_\sun]$	& 0.6958& 0.0010& 0.6355& 0.0007\\ 
$R$ [R$_\sun]$	& 0.687 & 0.020 & 0.609 & 0.016 \\ 
$\log{g}$	& 4.607 & 0.025 & 4.673 & 0.023 \\
$v_{rot}$ [km/s]\tablefootmark{b}
		&  8.72 &  0.25 &  7.73 &  0.21 \\
$\log(L/L_\sun)$& -0.687& 0.045 & -0.908& 0.064 \\
$M_{bol}$ [mag]	& 6.47 & 0.11 & 7.02 & 0.16 \\
&\multicolumn{4}{c}{\it From and based on Hipparcos} \\
$d$ [pc]	& \multicolumn{4}{c}{$31.1 \pm 1.1$}\\
$M_V$ [mag]	& 7.06 & 0.08 & 7.91 & 0.09 \\
&\multicolumn{4}{c}{\it From JKTABSDIM} \\
$d$ [pc]	& \multicolumn{4}{c}{$32.4 \pm 1.6$}\\
$M_V$ [mag]	& 6.91 & 0.17 & 7.68 & 0.27 \\
\hline
\end{tabular}
\tablefoot{
\tablefoottext{a}{Combined ASAS and SuperWASP solutions;}
\tablefoottext{b}{assuming synchronization;}
}
\end{table}

For the light curve (LC) analysis we used the latest version (v28)
of the code JKTEBOP \citep{sou04a,sou04b}, 
which is based on the EBOP program \citep{pop81}.
On the basis of spectroscopic data we first found the mass ratio 
and ephemeris, which we included in the LC analysis. We found that the
orbital period found directly by JKTEBOP from the ASAS photometry is in 
agreement with the one from RVs, however with larger uncertainties, and leading 
to significantly worse orbital solution. The SuperWASP data constrain the period 
even worse. For JKTEBOP we also used flux ratios found in TODCOR, and  
the logarithmic limb darkening (LD) law with coefficients
interpolated from the tables of \citet{vHa96} for ASAS and \citet{pol06}
for SuperWASP. The gravity darkening coefficients and bolometric albedos were 
always kept fixed at the values appropriate for stars with convective envelopes 
($g = 0.32$, $A = 0.5$). As mentioned before, various tests performed on 
every data set did not show a significant eccentricity of the orbit of AK~For, 
thus $e$ was kept fixed to 0 in the analysis. We fitted for the sum of the
fractional radii $r_1+r_2$, their ratio $k$, orbital inclination $i$,
surface brightness ratios $J$, and brightness scales (out-of-eclipse 
magnitudes in each filter).

We run the task 9, which to calculate reliable errors uses the residual-shifts 
method \citep{sou08} to asses the importance of the correlated ``red'' noise, 
especially strong in the SuperWASP data \citep{sou11}. 
We have run several tests to check how the final model varies with various 
LD coefficients and ephemeris, but we did not notice a strong dependence. 
Other sources of errors contribute more, but to at least partially account 
for LD coefficients and ephemeris uncertainties, we let them to be perturbed
in the residual-shifts simulations. It is a known fact that for systems with
partial eclipses, as AK~For, the orbital inclination
is correlated with the radii-related parameters, especially their sum. 
In Figure~\ref{fig_inc_rr} we show the results of the JKTEBOP analysis on the 
$r_1+r_2$ vs. $i$, and $k=r_2/r_1$ vs. $i$ diagrams. We see that different 
data sets give similar values of inclination, but different radii, 
nevertheless in agreement with each other. The probable reason for this slight 
inconsistency is the fact that we used the same flux ratio for the 
two such different filters. However, as mentioned before, no significant change
of the flux ratio was noticed in the wavelengths corresponding to the SuperWASP
filter. These differences can also result from the activity and the 
location of spots (not included here).

As the result we adopted weighted averages of the values found
from the two data sets. We mark them in Figure~\ref{fig_inc_rr}, together
with the adopted 1$\sigma$ errors. The model LCs for ASAS and SuperWASP 
photometry are presented in Fig. \ref{fig_lc}. The resulting values of 
fractional radii $r_{1,2}$, the inclination and surface brightness ratios
are given in Table~\ref{tab_par}.

The absolute values of parameters were calculated with the JKTABSDIM
procedure, available together with JKTEBOP. We assumed $E(B-V) = 0.0$, 
$JHK$ photometry from 2MASS \citep{skr06}, and $[M/H]=0.0$ or $-0.5$ -- 
values closest to our results of the spectral analysis. In both cases the 
results were practically undistinguishable. Distances were calculated with 
various bolometric corrections for various filters \citep{bes98,flo96,gir02} 
and surface brightness-$T_{eff}$ relations from \citet{ker04}. As the 
result we adopted their average and the standard deviation as the uncertainty. 
The value of 32.4 $\pm$ 1.6 pc is in agreement with the distance from 
\textit{Hipparcos} \citep[31.1 $\pm$ 1.1 pc;][]{vLe07}, which proves that 
the flux ratio from TODCOR and the temperatures we obtained in SME are correct. 
The single absolute $V$ magnitudes are also in agreement with values calculated 
on the basis of the \textit{Hipparcos} parallax and flux ratio from TODCOR,
however the latter have smaller errors. The complete set of final parameters 
is presented in the Table \ref{tab_par}. For the consistency we show distance 
and absolute $V$ magnitudes obtained in two ways, but adopt the 
{\it Hipparcos}-based values as the final ones. In this way we have a complete set 
of absolute physical parameters of the system not based on any models or 
calibrations.

\section{Discussion}

\subsection{Activity}

   \begin{figure}
   \centering
   \includegraphics[width=0.8\columnwidth]{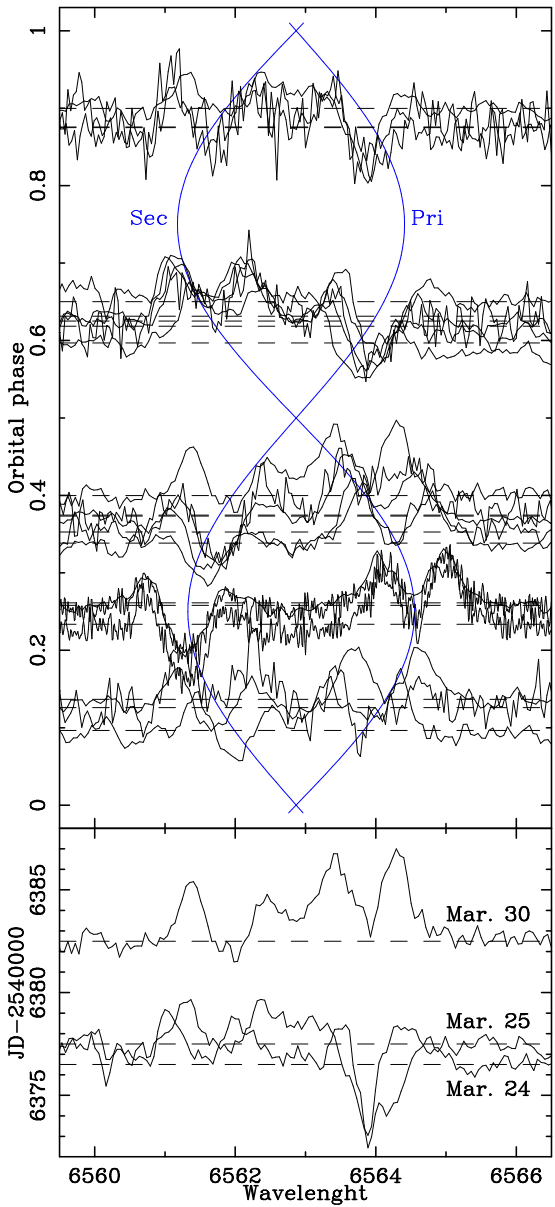}
   \caption{\ha emission lines of AK~For. In a given panel, all spectra were 
continuum-fitted and equally scaled. The dashed lines show 
the level of continuum and mark the orbital phase or time of the observation.
{\it Top}: All spectra are shown as a function of the orbital phase.
Blue solid lines trace the position of the \ha line as predicted by 
the orbital solution.
{\it Bottom}: Three FEROS spectra from March 2013, showing rapid evolution 
of the \ha profiles. In March 30 the emission was much stronger than few days
earlier. Secondary's emission is the stronger one.
}\label{fig_ha}
   \end{figure}

   \begin{figure}
   \centering
   \includegraphics[width=0.9\columnwidth]{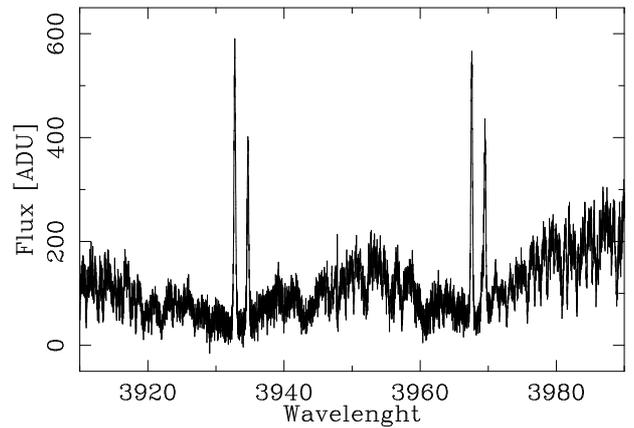}
   \caption{Portion of HARPS spectrum of AK~For around Ca~II~H and K lines.
Emission from both components in the cores is clearly visible. The 
apparently stronger lines come from the primary however, referred to the 
fainter continuum of the secondary star, its emission lines are intrinsically 
the strongest. Spectrum taken in quadrature.}\label{fig_ca}
   \end{figure}

   \begin{figure}
   \centering
   \includegraphics[width=0.9\columnwidth]{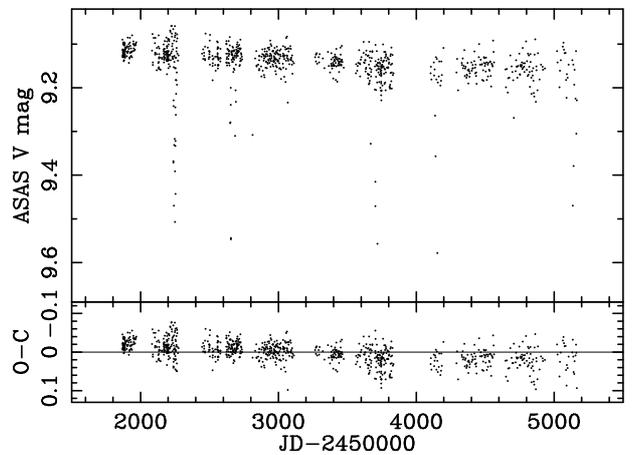}
   \caption{Long cadence ASAS light curve (top) and the residuals of the 
best-fitting model from the Figure \ref{fig_lc} (bottom) as a function of 
time. The change of the total brightness of the system is clearly visible, 
and can be explained by evolution of the spot pattern.}\label{fig_as_sp}
   \end{figure}

   \begin{figure}
   \centering
   \includegraphics[width=\columnwidth]{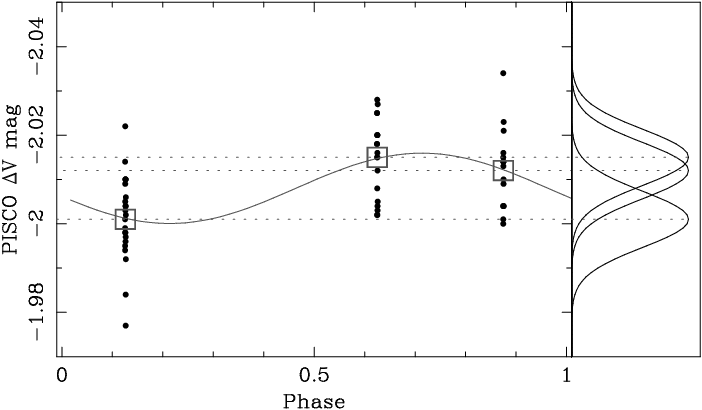}
   \caption{PISCO $V$-band photometry of AK For, taken simultaneously with 
CORALIE observations in February 2013, phase-folded with the orbital 
period. For each night squares mark the
average values, their error is 2 mmag (size of symbols) and 
the dispersion 
is 9 mmag, corresponding gaussians are shown to the right.
The presence of spots can be deduced from a small but clear fading around phase 0.126,
when presumably the spot is well visible. The brightness drop is 0.011 mag
relatively to the previous night (phase 0.875). The solid line is the sine fitted to data, 
plotted to guide the eye only and with no physical meaning.}\label{fig_pi_sp}
   \end{figure}
 
The activity of AK~For is undoubted, and was already noted \citep{gra06}. 
The system has been detected in the X-rays by the {\it Einstein} mission 
\citep{gio90}. We found the emission in \ha (Fig. \ref{fig_ha}) and, 
particularly strong, in the Ca~II H and K lines (Fig. \ref{fig_ca}; 
covered in our data only by HARPS). Both \ha lines have an interesting 
double-peak profile, with core absorption and emission in wings. 
The Ca~II lines of the primary appear more prominent, but taking into account the
flux ratio of two components around 3950 \AA, it is the secondary's
lines that are stronger with respect to the component's continuum.
The secondary also has stronger $H_\alpha$, but the \ha emission varies 
in the time scale of single days for both components (Fig. \ref{fig_ha}). 

The long-cadence ASAS photometry (Fig. \ref{fig_as_sp})
shows a clear fading trend by about 0.02 mag until
JD$\sim$2453500, probably originating in the evolution of spots.
The additional short-cadence photometry we made with PISCO 
(Fig. \ref{fig_pi_sp}) also proves the presence of spots,
causing brightness variations of a similar scale as the long-term
trend seen in the ASAS data. It is impossible however to constrain
the spot pattern, or even to point out the component on which they
are present -- most likely on both. If AK~For's brightness is still
at the level from September 2009 (the end of ASAS data), both 
components are probably substantially and uniformly covered with spots.
Additional, continuous, high-precision photometry from a global 
telescope network is required to constrain the pattern and location
of spots in a given time. With such data it would be also possible
to significantly improve the precision in radii, and make AK~For one
of the best-studied eclipsing systems.

\subsection{Kinematics}
The main sequence evolutionary stage of AK~For is suggested
by its galactic kinematics. We used our determinations of the systemic 
velocity together with the position, proper motion and distance
from {\it Hipparcos} (Tab. \ref{tab_lit}). The obtained 
values of $U = 17.8\pm0.9$~km/s, $V = -7.5\pm0.7$~km/s and
$W = 25.4\pm0.7$~km/s put marginally AK~For in the galactic thin disk
\citep{sea07}. This suggests the age below 4.5-5~Gyr, however 
older stars are observed in the thin disk \citep[][ and references therein]{all10}.
Taking into account the large value of $W$, at the edge of the thin
disk distribution, a young age (below 1 Gyr) seems to be unlikely.

\subsection{Comparison with theoretical models}

   \begin{figure}
   \centering
   \includegraphics[width=\columnwidth]{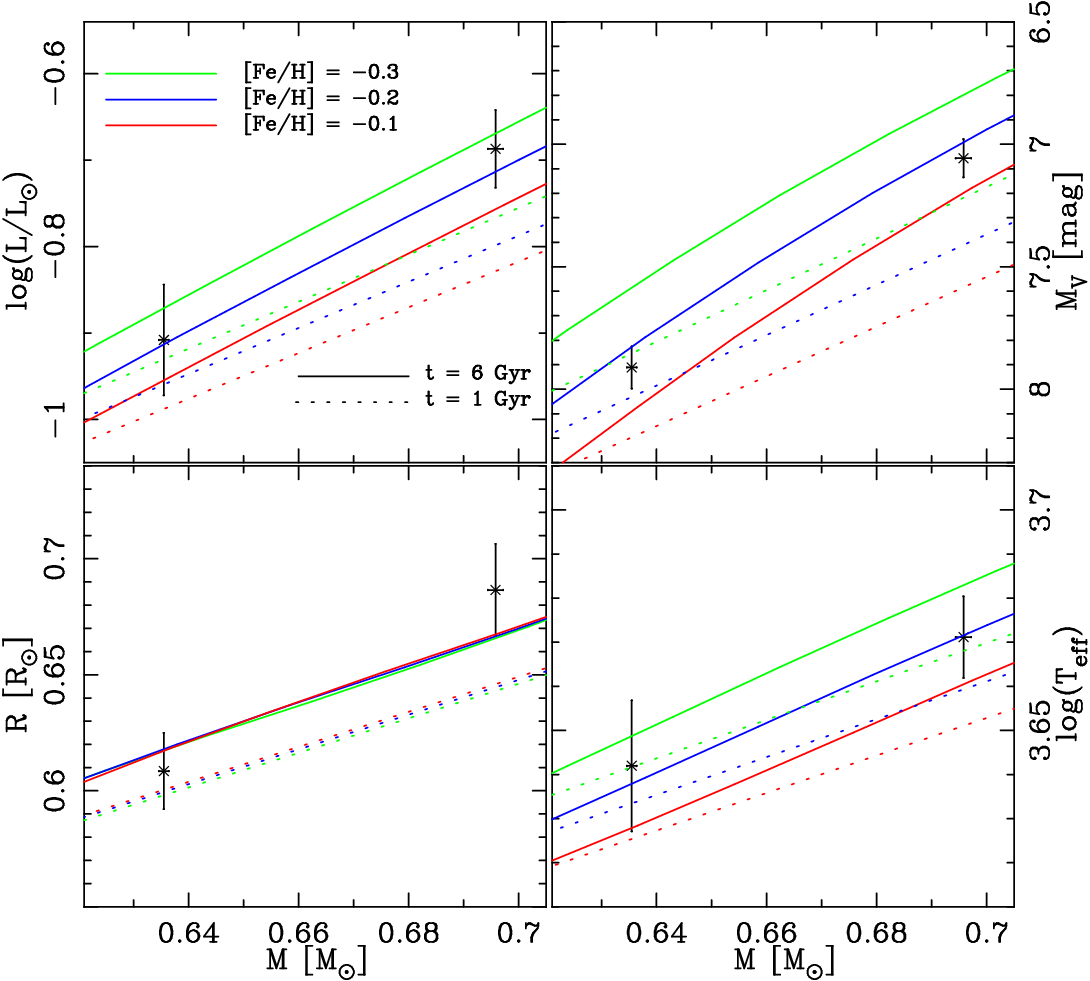}
   \caption{Comparison of our results with the Dartmouth stellar evolution models 
for three values of $[Fe/H]$, and ages of 1 and 6 Gyr.
}\label{fig_iso}
   \end{figure}

We reached a very high precision of 0.14+0.11\%
in the masses and also a good precision in radii (2.9+2.7\%),
allowing for meaningful tests of the stellar evolution models. 
AK~For has the most precise mass measurements among low-mass eclipsing binaries,
and it is the brightest one ($V$ = 9.135 mag from the ASAS light curve).
It is one of the most accurately measured eclipsing binaries in general.
The precision in masses was possible due to relatively low rotational 
velocities, related to the period longer than for the majority 
of similar DEBs. Improving the accuracy of the radii and related 
radiative parameters will require more precise and better-calibrated 
photometry.

In Fig.\,\ref{fig_iso} we 
compare our measurements with the Dartmouth stellar evolution 
models \citep{dot07} in the mass~$M$ vs. $R$, $\log(T_{eff})$, $\log(L/L_\sun)$ 
and $M_V$ planes. We show 1 and 6~Gyr isochrones for 
three values of $[Fe/H]$ spanning from -0.1 to 
-0.3~dex\footnote{SME gives $[M/H]$ but in case of 
the Dartmouth set the transformation to $[Fe/H]$ gives 
$-0.194^{+0.097}_{-0.099}$, thus we adopted $[Fe/H]=-0.2\pm0.1$}.

One can see that the 6~Gyr isochrone for the $[Fe/H]$ value obtained 
in the SME analysis fits to all the parameters within 1$\sigma$. The
system still resides on the main sequence, but has evolved and the 1~Gyr
models clearly do not fit the data. This shows 
how useful is the spectral analysis in solving the age-metallicity degeneracy
on the main sequence. The age of 6~Gyr was found to be the best-fitting one,
especially on the $M/R$ plane. It is in agreement with the kinematic 
analysis described in the previous section.

\subsection{0.4 -- 0.9 M$_\odot$ stars in DEBs}


AK Fornacis is an unique system for several reasons. First, it is 
bright and nearby, allowing for further detailed studies, which may however be 
problematic due to the unfortunate orbital period. Second, its 
components masses fall in the relatively poorly studied range of
0.5-0.8~M$_\odot$ -- between usually very active and mostly
convective M-dwarfs, and mostly inactive solar analogues of the G type.
The activity is thought to be the major factor responsible for the
observed discrepancies between measurements and theory in binaries, 
which tend to be larger for later spectral types. It is notable
that in this case the theory predicts the measurements reasonably well.
Finally, the orbital period of AK~For is longer than for the majority
of well-measured DEBs with K, M and late-G type components. 

In the online DEBCat catalogue\footnote{\texttt{http://www.astro.keele.ac.uk/$\sim$jkt/debcat/}
of well-measured eclipsing binaries, 
there are almost no stars of masses lower than 0.7~M$_\odot$
for periods longer than 3 days.} The notable exceptions are 
the secondary of \object{M55-V54} \citep[0.56~M$_\odot$, 9.27~d;][]{kal14},
the secondary of \object{KIC~6131659} \citep[0.69~M$_\odot$, 17.53~d;][]{bas12},
both components of \object{LSPM~J1112+7626} \citep[much lower masses 0.39+0.27~M$_\odot$, 41.03~d;][]{irw11}
or both components of \object{T-Lyr1-17236} \citep[0.68+0.53~M$_\odot$, 
8.43~d;][not included due to poor precision in radii]{dev08}.
It is obviously an observational selection effect and it is desirable 
to discover more long-period ($>3$~d) low-mass eclipsing binaries.

   \begin{figure}
   \centering
   \includegraphics[width=\columnwidth]{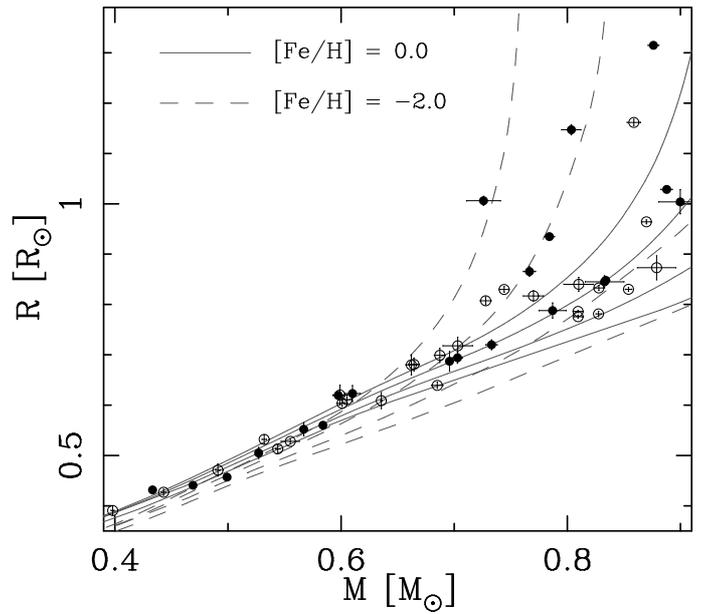}
   \caption{Mass-radius diagram for well-measured DEBs with at least one
component in the mass range of 0.4-0.9 M$_\odot$. Solid symbols represent
primary (more massive) and open symbols secondary components. 
Dartmouth isochrones of solar composition (solid) and
metal-depletion (dashed) are plotted for ages 1, 5, 10 and 14 Gyr.
}\label{fig_mr}
   \end{figure}

Fig.\,\ref{fig_mr} shows the mass-radius diagram for the well-measured 
eclipsing binaries with at least one component in the 0.4~--~0.9~M$_\odot$
mass range. We show systems from the DEBCat 
and three examples from our previous research \citep{hel11a,hel11b}.
Dartmouth isochrones for 1, 5, 10, 
14~Gyr and two metallicities are plotted over. It is clearly seen that 
age and metallicity plays an important role in the $M-R$ distribution, 
at least for stars more massive than 0.6~M$_\odot$. Traditionally, the radii and
temperatures of low mass stars were compared to the 1 Gyr, solar 
composition isochrone. The general characteristic was that the observed radii are 
larger and effective temperatures lower than models predicted. We now know a number of 
late type systems old enough to have their radii enlarged by the evolution 
(AK~For should be considered one of them). In some cases metallicity can also
be estimated from the spectra, which helps to constrain the age.
As one can see, the traditional approach is not always valid. 
In order to indicate if a given star's radius is enlarged or not, one needs
an independent metallicity and age estimation (possible only in clusters), 
or may check if the slope that the two components define on the $M-R$ diagram
is compatible with any normal evolutionary relation.
Usually the secondary appears to be oversized (over-aged), but
literature examples show that it may also be the primary.

According to the recently popular explanation of the 
observed oversized radii of low-mass DEBs, the fast rotational velocity, 
resulting from short orbital period, is responsible for higher 
level of activity, which inhibits the effectiveness
of convection, leading to larger radii and 
lower temperatures \citep{cha07}. Long-period systems should thus
be less active and have their radii and temperatures better 
reproduced than the short-period binaries. One can see in Fig.\,\ref{fig_iso} 
that AK~For does not follow this trend, despite being active.
On the other hand, the 0.85~M$_\odot$ secondary of a 4.8~d system 
\object{V636~Cen} is both active and inflated \citep{cla09}, as well
as both components of the 41~d period \object{LSPM~J1112+7626} \citep{irw11}.
It is worth mentioning that the Ca~II emission lines are much stronger in 
AK~For than in V636~Cen, which rotates a bit faster. Some of the 
components of longer-period systems found in globular clusters, like 
the secondaries of V66 and V69 in \object{M4} \citep{kal13} 
or the primary of V54 in \object{M55} \citep{kal14}, also seem to 
be oversized, but their activity is not seen. In the same time we 
also know short-period active systems, like \object{ASAS~J045304-0700.4} 
\citep[$P=1.77$~d;][]{hel11a}, \object{MG1-78457} ($P=1.59$~d), 
\object{MG1-506664} \citep[$P=1.55$~d;][]{kra11}, or 
\object{KOI-126 BC} \citep[$P=1.77$~d;][]{fei11}, where 
the model radii fit the observed ones. In the light of recent findings 
it is difficult to draw any conclusions about the connection
between rotation, activity, radii and effective temperatures.

\section{Summary}
We present the first full orbital and physical analysis of the brightest 
low-mass detached eclipsing binary known to date -- AK~Fornacis.
The system's characteristics (brightness, long orbital period) allow for
a very accurate determination of masses, and fairly good of the radii.
The precision in masses is very high but still somewhat limited 
by the stellar activity. The object's characteristics are unusual among
the known low-mass DEBs, as despite being very active, properties of
both components are nicely reproduced and the age can be 
determined fairly well as for a main sequence system. The reason for the
enhanced activity can not be the rotation, as AK~For rotates rather
slowly, although faster than typical, single field K and M dwarfs.

The major observational disadvantage of the 
system is its period, very close to 4~d, which makes the eclipses not 
always possible to observe from one place during a single season. 
Further work on this system requires continuous photometry from a 
global network of telescopes, such as WET \citep{nat90}, 
{\it Solaris} \citep{kon12} or HAT-S \citep{bak09}, in order to derive 
some of the crucial parameters, like the radii or temperatures, even 
more accurately. 

\begin{acknowledgements}
We would like to thank Dr Johannnes Andersen for the discussion and valuable
comments that helped us improve our manuscript, 
and the staff of the La Silla and Geneva observatories, 
especially G. Lambert and D. Naef, for their hospitality 
and help during the observations.\\
K.G.H. acknowledges support provided by the Proyecto FONDECYT Postdoctoral 
No. 3120153, the Polish National Science Center grant 2011/03/N/ST9/01819,
and the National Astronomical Observatory of Japan as Subaru Astronomical 
Research Fellow.
R.B. and N.E. are supported by CONICYT-PCHA/Doctorado Nacional.
M. Ratajczak is supported by the Polish National Science Center through grant 2011/01/N/ST9/02209.
A.J. acknowledges support from FONDECYT
project 1130857, BASAL CATA PFB-06, and grant IC120009 awarded to the
Millennium Institute of Astrophysics, MAS, by the Millennium Science
Initiative of the Chilean Ministry of Economy, Development, and
Tourism.
M.K. is supported by the European Research Council Starting Grant, the
Polish National Science Center through grant 5813/B/H03/2011/40, the
Ministry of Science and Higher Education through grant W103/ERC/2011 
and the Foundation for Polish Science through grant "Ideas for Poland".
M. Rabus acknowledges support from FONDECYT postdoctoral fellowship No. 3120097.\\
We have used data from the WASP public archive in this research. 
The WASP consortium comprises 
of the University of Cambridge, Keele University, University of 
Leicester, The Open University, The Queen’s University Belfast, 
St. Andrews University and the Isaac Newton Group. Funding for 
WASP comes from the consortium universities and from the UK’s 
Science and Technology Facilities Council. 
This publication makes use of data products from the 
Two Micron All Sky Survey, which is a joint project of the 
University of Massachusetts and the Infrared Processing and 
Analysis Center/California Institute of Technology, funded by 
the National Aeronautics and Space Administration and the 
National Science Foundation. 
\end{acknowledgements}

\bibliographystyle{aa}


\end{document}